\documentclass[twocolumn,showpacs,preprintnumbers]{revtex4}
\usepackage{amssymb}
\usepackage{amsfonts}
\usepackage{amsmath}
\usepackage{graphicx}
\usepackage{dcolumn}
\usepackage{bm}

\input{tcilatex}

\begin{document}

\title{On the Phase Boundaries of the Integer Quantum Hall Effect. II}
\author{S. S. Murzin}
\affiliation{Institute of Solid State Physics RAS, 142432, Chernogolovka, Moscow
District, Russia }

\begin{abstract}
It is shown that the statements about the observation of the transitions
between the insulating phase and the integer quantum Hall effect phases with
the quantized Hall conductivity $\sigma _{xy}^{q}$ $\geq 3e^{2}/h$ made in a
number of works are unjustified. In these works, the crossing points of the
magnetic field dependences of the diagonal resistivity $\rho _{xx}$ at
different temperatures $T$ at $\omega _{c}\tau \approx 1$ have been
misidentified as the critical points of the phase transitions. In fact,
these crossing points are due to the sign change of the derivative $d\rho
_{xx}/dT$ owing to the quantum corrections to the conductivity. Here, $%
\omega _{c}$ is the cyclotron frequency, $\tau $ is the transport relaxation
time.
\end{abstract}

\maketitle

\narrowtext

The phase diagram of two dimensional systems in the magnetic field has
attracted the attention of both theorists and experimentalists for many
years. Treating the integer quantum Hall effect (IQHE) in the context of the
two parameter scaling theory [1], which is graphically represented as a flow
diagram [2, 3], yields the solution of the problem disregarding the
electron--electron interaction. The further development of the scaling
theory showed that the electron--electron interaction does not affect the
position of the IQHE phase boundaries of a spin polarized electron system
[4, 5].

According to the scaling theory, the boundary between two IQHE phases is
possible only if the quantized values of the Hall conductivity of these two
phases differ by $e^{2}/h$ or (in the case of the spin degeneracy of the
Landau levels) $2e^{2}/h$. However, a number of works reported on the
observation of the transitions between the insulating phase ($\sigma
_{xy}^{q}=0$) and the IQHE phases with the quantized Hall conductivity $%
\sigma _{xy}^{q}\geq 3e^{2}/h$. Song et al. [6] reported on the observation
of the transition $\sigma _{xy}^{q}=0\leftrightarrow \sigma
_{xy}^{q}=3e^{2}/h$ in two dimensional hole systems in a strained Ge quantum
well. The observation of the transitions $\sigma _{xy}^{q}=0\leftrightarrow
\sigma _{xy}^{q}\geq 3e^{2}/h$ in the two dimensional hole systems in a
strained Ge quantum well was also announced in [7]. Lee et al. [8] claimed
the observation of the transitions $\sigma _{xy}^{q}=0\leftrightarrow \sigma
_{xy}^{q}=6e^{2}/h$ and $\sigma _{xy}^{q}=0\leftrightarrow \sigma
_{xy}^{q}=8e^{2}/h$ in doped AlGaAs/GaAs/AlGaAs quantum wells. Huang et al.
[9] reported on the observation of the transitions from the state with $%
\sigma _{xy}^{q}=0$ to the states with $\sigma _{xy}^{q}=6-16e^{2}/h$ in
GaAs/AlGaAs heterojunctions. In all of these works [6--9], the crossing
points of the magnetic field dependences of the diagonal resistivity $\rho
_{xx}$ at different temperatures $T$, at $\omega _{c}\tau \approx 1$\ ($%
\omega _{c}=eB/m$ is the cyclotron frequency, $\tau $ is the transport
relaxation time, and $m$ is the effective electron mass) were considered as
the critical points $B_{c}$ of the phase transitions (see Fig. 1). In this
case, $\rho _{xx}$ weakly depends on the magnetic field and temperature near 
$B_{c}$.

\begin{figure}[t]
\includegraphics[width=8.5cm,clip]{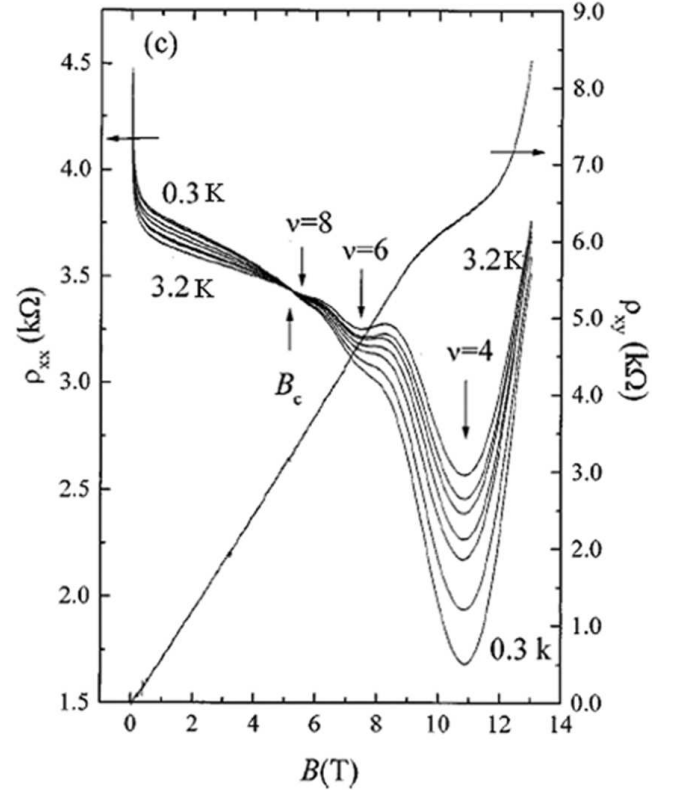}
\caption{The diagonal ($\protect\rho _{xx}$) and Hall ($\protect\rho _{xy}$%
)\ resistivities of the doped AlGaAs/GaAs/AlGaAs quantum well as a function
of the magnetic field. The electron density is $N_{s}$ $=1.04\times 10^{16}$
m$^{-2}$ [8]. The temperatures for different $\protect\rho _{xx}$- curves
are 0.3, 0.5, 0.8, 1.2, 1.7, 2, and 3.2 K. The spin splitting is small and
therefore invisible in $\protect\rho _{xx}$ and $\protect\rho _{xy}$. The
figure is taken from Ref..}
\label{R}
\end{figure}

In this work, it is shown that the above statements of the observation of
the transitions between the insulating phase and the IQHE phases with the
quantized Hall conductivity $\sigma _{xy}^{q}\geq 3e^{2}/h$ are unjustified.
In fact, the crossing point of the magnetic field dependences of the
diagonal resistivity $\rho _{xx}(T)$ at different temperatures is caused by
the sign change of the derivative $d\rho _{xx}/dT$ owing to the quantum
corrections to the conductivity [10]. 
\begin{figure}[t]
\includegraphics[width=8.5cm,clip]{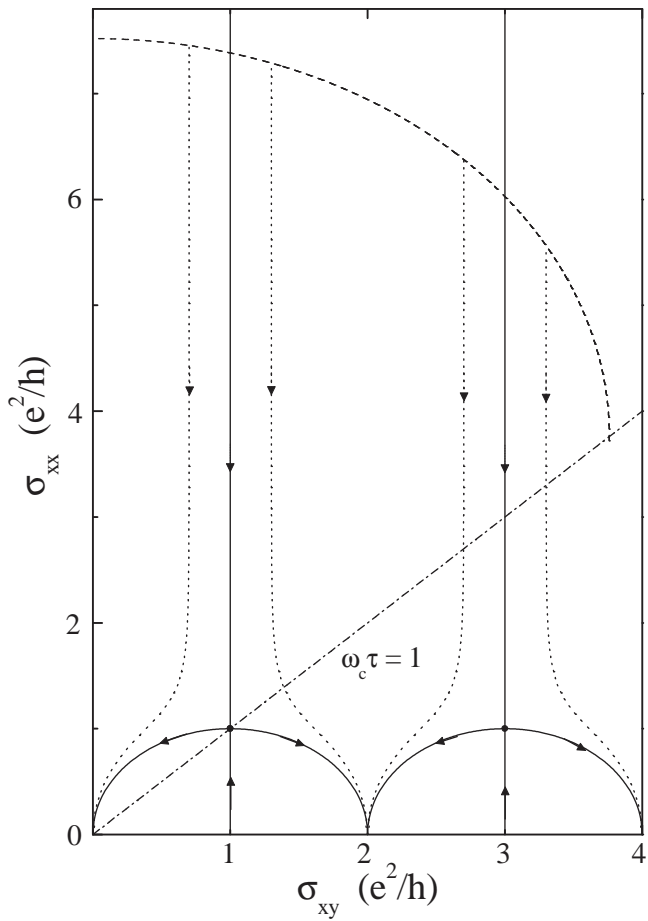}
\caption{Sketch of the scaling flow diagram for the quantum well with the
parameters given in Fig. 1. The spin splitting is negligible. The solid
lines are the separatrices of the diagram. The dashed line shows dependence $%
\protect\sigma _{xx}(\protect\sigma _{xy})$ at $\protect\omega _{c}\protect%
\tau <1$ for the sample with the zero-field bare conductivity $\protect%
\sigma ^{0}=7.52e^{2}/h$. The dotted lines are the scaling flow lines. The
dash--dotted straight line corresponds to $\protect\omega _{c}\protect\tau %
=1 $.}
\end{figure}

The classical diagonal and Hall conductivities in the magnetic field take
the form 
\begin{equation}
\sigma _{xx}^{0}=\frac{N_{s}e^{2}\tau }{m}\frac{1}{1+\left( \omega _{c}\tau
\right) ^{2}}  \label{Gxx}
\end{equation}%
and 
\begin{equation}
\sigma _{xy}^{0}=\frac{N_{s}e^{2}\tau }{m}\frac{\omega _{c}\tau }{1+\left(
\omega _{c}\tau \right) ^{2}}.  \label{Gxy}
\end{equation}%
where $N_{s}$ is the electron or hole density. The quantum corrections to
the diagonal conductivity $\Delta \sigma _{xx}(T)=\sigma _{xx}(T)-\sigma
_{xx}^{0}\ll \sigma _{xx}^{0}$ decrease with temperature. At $T\rightarrow 0$%
, $\sigma _{xx}(T)\rightarrow 0$ except for the critical points, where $%
\sigma _{xy}^{0}=(i+1/2)e^{2}/h$ (see Fig. 2). Excluding the weak
localization region ($B\lesssim 1$ T), the Hall conductivity $\sigma _{xy}$
is independent of the temperature down to the temperatures at which $\sigma
_{xx}\sim e^{2}/h$. At lower temperatures, the Hall conductivity depends on
the temperature and approaches the nearest quantized integer value $\sigma
_{xy}(B_{i})=ie^{2}/h$ with $B_{i}<B_{c}$ at $\omega _{c}\tau <1$.

Taking into account the quantum corrections, the diagonal and Hall
resistivities of the two dimensional electron system are given by the
expressions 
\begin{equation}
\rho _{xx}(T)=\rho _{xx}^{0}+\left[ \left( \rho _{xy}^{0}\right) ^{2}-\left(
\rho _{xx}^{0}\right) ^{2}\right] \Delta \sigma _{xx}(T)  \label{R1}
\end{equation}%
and 
\begin{equation}
\rho _{xy}(T)=\rho _{xy}^{0}-2\rho _{xx}^{0}\rho _{xy}^{0}\Delta \sigma
_{xx}(T)  \label{R2}
\end{equation}

Here, we took into account that $\rho _{xx}^{0}=\sigma _{xx}^{0}/\left[
\left( \sigma _{xx}^{0}\right) ^{2}+\left( \sigma _{xy}^{0}\right) ^{2}%
\right] $, $\rho _{xy}^{0}=\sigma _{xy}^{0}/\left[ \left( \sigma
_{xx}^{0}\right) ^{2}+\left( \sigma _{xy}^{0}\right) ^{2}\right] $ and $%
\sigma _{xx}^{0}$, $\sigma _{xy}^{0}$, $\rho _{xx}^{0}$ and $\rho _{xy}^{0}$
are the bare (non-renormalized) values of the conductivity and resistivity,
which correspond to the diffusion motion of electrons without the
interference (localization) effects at distances longer than the diffusion
step length. The derivative $d\rho _{xx}/dT$ changes its sign in the
magnetic field $B$ such that $\rho _{xx}^{0}(B)=\rho _{xy}^{0}(B)$. In the
classical treatment $\rho _{xx}^{0}(B)=\rho _{xy}^{0}(B)$ at $\omega
_{c}\tau =1$.

At $\sigma _{xy}\gg e^{2}/h$ and $\omega _{c}\tau <1$, the diagonal
resistivity $\rho _{xx}$ first increases with a decrease in the temperature,
reaching the value 
\begin{equation}
\rho _{xx,\max }=\frac{1}{2\sigma _{xy}^{0}},  \label{max}
\end{equation}%
and then decreases and vanishes at $T\rightarrow 0$ excluding the critical
magnetic fields in which $\sigma _{xy}^{0}=(i+1/2)e^{2}/h$. Thus, the
negative value of the derivative $d\rho _{xx}/dT$ within the experimental
range does not imply that the electron system is in an insulating phase.

Note that the magnetic field position of the IQHE phases at $\omega _{c}\tau
\lesssim 1$ is not determined by the filling factor $\nu $. Rather, it is
given by the magnitude of $\sigma _{xy}^{0}h/e^{2}$. This quantity is
different from $\nu $ at $\omega _{c}\tau \lesssim 1$. [11] At $\nu =8$ in
Fig.1 $\sigma _{xy}^{0}h/e^{2}=3.8$, $\sigma _{xx}^{0}h/e^{2}=3.8$ at $T=0.3$
K. According scaling diagram presented on Fig2 at this $\nu $ quantized
values of the Hall conductivity $\sigma _{xy}^{q}=4e^{2}/h$.

Thus, we have shown that the statements [6--9] of the observation of the
transitions between the insulating phase and the IQHE phases with the
quantized Hall conductivity $\sigma _{xy}^{q}\geq 3$ are unjustified. In
fact, the crossing point of the magnetic field dependences of the diagonal
resistivity $\rho _{xx}(T)$ at different temperatures is caused by the sign
change of the derivative $d\rho _{xx}/dT$ at $\omega _{c}\tau \approx 1$
owing to the quantum corrections to the conductivity.

This work was supported by the Russian Foundation for Basic Research.

\end{document}